# Observation of Space-Charge-Limited Transport in InAs Nanowires

A. M. Katzenmeyer, F. Léonard, A. A. Talin, M. E. Toimil-Molares, J. G. Cederberg, J. Y. Huang and J. L. Lensch-Falk

*Abstract*—Recent theory and experiment have suggested that space-charge-limited transport should be prevalent in high aspect-ratio semiconducting nanowires. We report on InAs nanowires exhibiting this mode of transport and utilize the underlying theory to determine the mobility and effective carrier concentration of individual nanowires, both of which are found to be diameter-dependent. Intentionally induced failure by Joule heating supports the notion of space-charge limited transport and proposes reduced thermal conductivity due to the nanowires' polymorphism.

*Index Terms*—Carrier concentration, InAs, Mobility, Nanowires, Space-charge-limited.

## I. Introduction

Semiconductor nanowires (NWs) continue to fascinate researchers, promising control over dimensions, crystallinity and composition during the nanostructure synthesis, as well as interesting basic transport properties and potential applications in electronic, optoelectronic, and thermoelectric devices. In order for these nanostructures to have technological impact, their basic electrical characteristics have to be measured accurately and reproducibly; however, bulk techniques such as Hall measurements cannot be easily implemented with NWs due to the small dimensions. Characteristic nanowire parameters such as carrier type, concentration and mobility are thus frequently determined from the transfer characteristics of NW field-effect transistors (NWFETs) [2]. The results, however, can be strongly affected by the nanowire and the dielectric/nanowire surface and interface states [3]-[4]. For example, Dayeh et al. found that the transconductance in top-gated InAs NWFETs is dependent on the sweep-rate of the gate voltage and that the hysteresis in the transfer characteristic is only removed by creating a charge-neutral interface via a slow sweep-rate [5]. This naturally leads to a dependence of the extracted carrier concentration and mobility on the experimental conditions, with the extracted mobility varying by up to an order of magnitude. Thus, alternative characterization techniques free of perturbing interfaces would be valuable to measure the intrinsic properties of NWs.

Here, we report on the electrical characterization of VLS (vapor-liquid-solid) grown InAs NWs using such a technique. We contact individual, free-standing NWs directly on their growth substrate using an electronic nanoprobe retrofitted inside of a field-emission scanning electron microscope (FE-SEM) as shown in Fig. 1 (a). The tungsten (W) probe serves as one contact and the heavily doped growth substrate serves as the other. Using this technique, we show that the electrical transport in these InAs NWs is space-charge-limited (SCL), a bulk transport regime that arises when the charge injection at the contacts is efficient. Using recently developed theory specific to SCL transport in NWs [6], we show how one can extract the mobility and effective carrier concentration from I-V measurements indicative of SCL current. By controllably inducing failure by Joule heating, we show that electrical transport is dominated by the bulk of the NW instead of the contacts, a further indication of SCL transport. The two-point probe method we employ [7] is performed in-vacuo on free-standing wires thus diminishing the electrical and thermal influence of molecular adsorbates and the substrate on which NW devices typically lie.

## II. Nanowire Synthesis

Non-intentionally-doped InAs NWs were grown on a degenerately Si-doped GaAs(111)B substrate. The substrate was coated with a 1 nm thick Au layer by ex-situ electron beam evaporation, and subsequently inserted into the chemical vapor deposition (CVD) reactor. In the growth chamber, the template was annealed for 10 min in arsine ($AsH_3$) ambient at 650°C and then cooled to 400°C for growth. The NWs were grown by low-pressure (60 Torr) metal-organic CVD using

Manuscript received December 22, 2009.

This project was supported by the Laboratory Directed Research and Development program at Sandia National Laboratories, a multiprogram laboratory operated by Sandia Corporation, a Lockheed Martin Company, for the United States Department of Energy under contract DE-AC04-94-AL85000.

A. M. Katzenmeyer is with Sandia National Laboratories, Livermore, CA 94551 USA and the University of California at Davis, CA 95616 USA.

F. Léonard is with Sandia National Laboratories, Livermore, CA 94551 USA (phone: 925-294-3511; fax: 925-294-3231; e-mail: fleonar@sandia.gov).

A. A. Talin was with Sandia National Laboratories, Livermore, CA 94551 USA. He is now at the National Institute of Standards and Technology, Gaithersburg, MD 20899 USA

M. E. Toimil-Molares was with Sandia National Laboratories, Livermore, CA 94551 USA. She is now at GSI Helmholtzzentrum für Schwerionenforschung GmbH, Planckstrasse 1, D-64291 Darmstadt, Germany.

J. G. Cederberg is with Sandia National Laboratories, Albuquerque, NM 87185 USA.

J. Y. Huang is with Sandia National Laboratories, Albuquerque, NM 87185 USA.

J. Lensch-Falk is with Sandia National Laboratories, Livermore, CA 94551 USA.

AsH$_3$ and trimethyl indium (TMIn) with the Au serving as the collector for the VLS mechanism. The growth occurred for 10 min using a TMIn partial pressure of 2.42 mTorr and an AsH$_3$ partial pressure of 0.336 Torr giving a V-III ratio of 140. The resulting NWs have diameters between 30 and 300 nm and lengths often exceeding 10 µm. Planar growth rates under these conditions were below 0.04 µm/min. Transmission electron microscope (TEM) images [Fig. 1 (b)] and selective area electron diffraction (SAED) [Fig. 1 (c)] indicate that the wires are primarily wurtzite (WZ), are covered by ~2 nm native oxide, and contain a number of stacking faults (SFs) and thin zinc-blende (ZB) regions. The WZ structure with SFs and ZB regions is commonly observed in III-V NWs grown on 111(B) substrates. Growth conditions and the VLS mechanism make it energetically favorable for the WZ phase to occur [8].

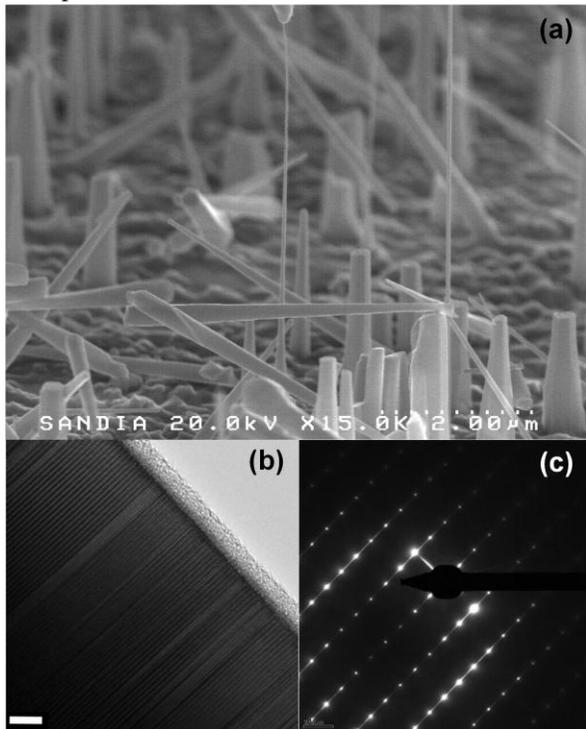

Fig. 1 (a) FE-SEM micrograph depicting a typical I-V measurement. In the center of the image, a W nanoprobe contacts a free-standing NW at its tip. The degenerately-doped substrate serves as the second electrode. (b) TEM image indicating WZ (dark) and ZB (light) regions (scale bar is 5 nm). (c) Electron diffraction pattern of a typical WZ region.

## III. SPACE-CHARGE-LIMITED TRANSPORT

Fig. 2(a) displays representative I-V characteristics for three NWs of different dimensions. The curves are non-linear, with symmetric behavior between positive and negative voltage polarities. Such I-V curves are often reported in the characterization of NWs using lithographically-defined contacts, and are usually attributed to the presence of symmetric back-to-back Schottky or tunnel barriers at the contacts. However, in the present measurements the contacts are very dissimilar, and yet the curves are symmetric. Moreover, plots of I/V vs V show that in addition to the small bias Ohmic regime, there is a strong quadratic $I \propto V^2$ dependence on voltage, as shown in Fig. 2(b), and we attribute this behavior to space-charge limited (SCL) transport as was recently modeled and measured for GaN NWs [6]. SCL transport can be dramatically enhanced in NWs due to poor electrostatic screening. That is, the current density $J$ is given by

$$J = \varsigma\left(\frac{R}{L}\right)\frac{\varepsilon\mu}{L^3}V^2 \quad (1)$$

where $\varsigma(R/L)$ is a scaling factor which depends on the aspect ratio of the NW. In the limit $R/L \ll 1$ appropriate for the NWs included in this study,

$$\varsigma\left(\frac{R}{L}\right) = \varsigma_0\left(\frac{R}{L}\right)^{-2} \quad (2)$$

where $\varsigma_0$ is a numerical constant of order unity [9]. For high aspect-ratio wires, the scaling factor induces a strong deviation from the bulk current density for which $\varsigma = 9/8$.

Solving for the mobility of a high aspect-ratio NW in the SCL transport regime $\mu_{SCL}$, gives

$$\mu_{SCL} = \frac{IL}{V^2\pi\varepsilon} \quad (3)$$

where $\varepsilon$ is the permittivity of InAs. The mobility can thus be extracted directly from the SCL regime of a 2 point probe I-V

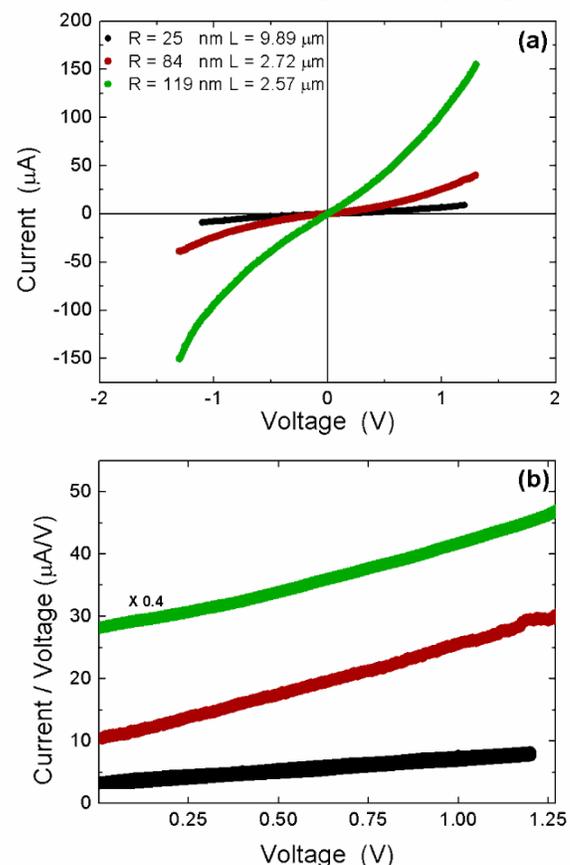

Fig. 2. (a) Current as a function of voltage for three representative NWs of different dimensions showing symmetric, non-linear, non-exponential behavior. (b) Plots of current/voltage as a function of voltage for the wires shown in (a).

measurement, in addition to the linear Ohmic regime at small bias. The cross-over voltage, $V_c$, where the Ohmic, $J = n_{eff} e \mu L^{-1} V$, and SCL current, Eq. (1), are equal allows extraction of the effective carrier concentration:

$$n_{eff} = \frac{V_c \varepsilon}{eR^2}. \quad (4)$$

Plotting $I/V$ as a function of $V$, as shown in Fig. 2(b), yields a straight line for SCL behavior and allows one to extract the Ohmic (y-intercept) and the SCL (slope) current contributions. The length and radius of the wires were determined from high resolution FE-SEM micrographs. Fig. 3(a) displays $n_{eff}$ as a function of NW radius indicating that the effective carrier concentration increases with decreasing diameter. The trend is well fit by a power law behavior which shows that $n_{eff} \propto R^{-2}$. In general, the effective carrier concentration is expected to increase with decreasing radius because the InAs surface is inverted with a high density of surface charge [10-12]. The thickness of the surface layer has been reported to be around $t \approx 10 nm$ [13], leading to a volume density on the order of $10^{18}$ cm$^{-3}$. As the nanowire radius decreases, this surface layer becomes more important, leading to an increase of the effective carrier concentration. From a surface to volume argument, this leads to a $n_{eff} \propto R^{-1}$ dependence; however, it is known that band-bending distances depend on nanowire radius [14] so we also expect $t$ to increase with decreasing nanowire radius. These two factors together may explain the rapid increase of the effective carrier concentration with decreasing radius that we observe experimentally. The mobility as a function of radius is given in Fig. 3(b), and shows an increase with increasing nanowire diameter. This trend in mobility as a function of radius is consistent with that observed by Ford et al. [15] in small diameter wires. In general, the values we find for mobility and carrier concentration are in agreement with the findings of other groups investigating transport in InAs NWs [16-18] including one study dealing specifically with WZ/ZB NWs similar to our own [19]. The main source of scatter in the extracted values is most likely due to the different stacking fault density and arrangement from wire to wire.

## IV. JOULE HEATING

Since the non-linear, symmetric I-V curves of Fig. 2(a) are often ascribed to contacts, we sought an additional test to show that the I-V curves were predominantly determined by the transport in the NW itself. Thus, we intentionally Joule-heated NWs to their maximum temperature (i.e. thermal breakdown) in order to determine the location of failure. For a contact-dominated situation, this point would be close to the contacts since this is where the greatest voltage drop would occur. In contrast, for transport dominated by the NW itself, this point should be close to the midpoint between the two contacts. Fig. 4(a) shows a SEM image of an InAs NW after breakdown, clearly indicating failure in the middle of the NW. This result confirms the presence of low resistance contacts [20] and that the transport characteristics are determined by the NW. We carried out similar in-situ failure experiments on a number of NWs of different dimensions, recording the current and voltage at breakdown. (A movie of one of the Joule failure experiments is available as supplementary downloadable material at http://ieeexplore.ieee.org.) The temperature profile in the NWs was estimated using a 1-dimensional heat flow model [21]. The model assumes that the temperature distribution in the NW depends only on the axial coordinate, a condition that is appropriate in our case given that the wires are freestanding in vacuum, with little heat flow to the environment. Under this condition, the temperature distribution is the same as that in a bulk material. For two ends held at temperature $T_0$, the temperature profile is given by

$$T(z) = T_0 + \frac{q}{2\kappa} z(L-z) \quad (5)$$

where $z$ is the axial coordinate, $q$ is the heat generation rate, $L$ is the length of the wire, and $\kappa$ is the thermal conductivity. Within this model, the maximum temperature is reached at the point $z = L/2$, i.e. at the midpoint between the two contacts, as observed in our experiments. By equating the heat generation rate with the electrical power delivered to the wire, the maximum temperature is

$$T_{max} = \frac{IVL}{8\kappa\pi R^2}. \quad (6)$$

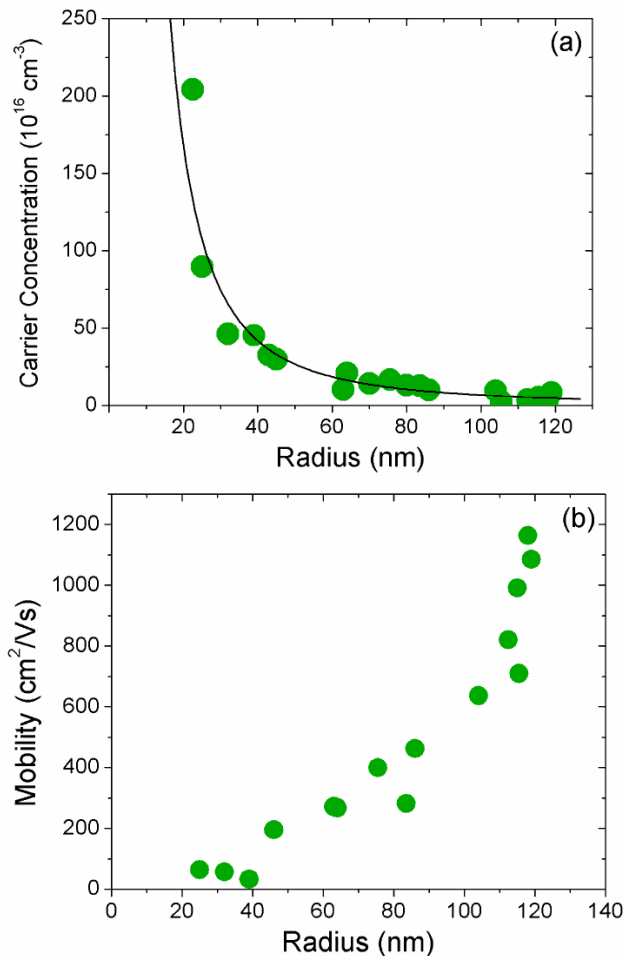

Fig. 3. (a) Effective carrier concentration and (b) mobility plotted as a function of radius. Carrier concentration depends inversely on NW cross-sectional area and mobility is proportional to the radius in thin wires.

We used this equation to extract the thermal conductivity of NWs of different radii and length by measuring the current and voltage at which the NWs breakdown, and equating $T_{max}$ to 793 K, a temperature known to result in the desorption of the native oxide [22-23]. Assuming that the NW failure is initiated at 793 K [24-25], the thermal conductivity is plotted for NWs as a function of their radius in Fig. 4 (b). The average thermal conductivity extracted from the failure measurements is 8.35 W/mK, about 1/3$^{rd}$ the value of high quality bulk InAs. This result is in accord with the large density of stacking faults present in these NWs, as it is well-known that structural defects, including stacking faults, cause increased phonon scattering [26].

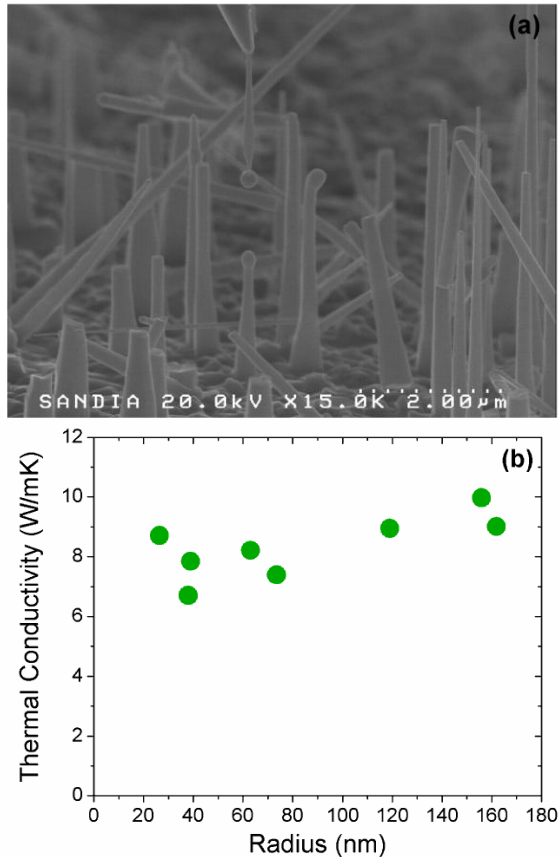

Fig. 4. Panel (a) shows a SEM image of an InAs NW after breakdown by Joule heating. (b) Thermal conductivity as a function of NW radius extracted from the breakdown experiments.

## V. CONCLUSION

We have presented a two-point probe method for electrical characterization of NWs which does not rely on the FET geometry. When applied to InAs NWs, the measurements suggest that space-charge-limited current is the dominant transport mechanism. For our InAs NWs an inverse relationship was found between carrier concentration and NW cross-sectional area while carrier mobility was found to increase with NW radius. Thermal breakdown measurements suggest a suppression of the thermal conductivity with respect to bulk which may be a result of numerous stacking faults along the length of the wires. These results indicate that dimensionality effects govern the electrical behavior of InAs NWs even at radii as large as 100 nm.